\documentclass[]{spie}
\usepackage[]{graphicx}

\def\bicep{{\sc Bicep}}

\def\dasi{{\sc Dasi}}

\def\biceptwo{{\sc Bicep2}}

\def\planck{{\it Planck}}
\def\quiet{{\sc Quiet}}
\def\spider{{\sc Spider}}

\def\QUAD{{\sc QUaD}}
\def\cbi{{\sc CBI}}

\def\ebex{{\sc EBEX}}
\def\boom{{\sc Boomerang}}
\def\wmap{{\sc Wmap}}

\def\keck{{\sc Keck}}
\def\maxipol{{\sc Maxipol}}
\def\polarbear{{\sc PolarBear}}
\def\iram{{\sc IRAM}}
\def\epic{{\sc EPIC}}

\def\synfast{{\tt synfast}}
\def\healpix{{\tt Healpix}}

\def\muK{~\mu{\rm K}}

\def\deg{^\circ}

\title{Absolute polarization angle calibration using polarized diffuse Galactic emission observed by BICEP}
\vspace{-0.5cm}
\author{Tomotake Matsumura\supit{a},
Peter Ade\supit{b}, 
Denis Barkats\supit{c}, 
Darcy Barron\supit{d}, 
John O. Battle\supit{e}, 
Evan M. Bierman\supit{e}, 
James J. Bock\supit{e,a}, 
H. Cynthia Chiang\supit{f}, 
Brendan P. Crill\supit{e,a},
C. Darren Dowell\supit{e,a}, 
Lionel Duband\supit{g}, 
Eric F. Hivon\supit{h}, 
William L. Holzapfel\supit{i}, 
Viktor V. Hristov\supit{a}, 
William C. Jones\supit{f}, 
Brian G. Keating\supit{e}, 
John M. Kovac\supit{j}, 
Chao-Lin Kuo\supit{k}, 
Andrew E. Lange\supit{e,a}, 
Erik M. Leitch\supit{e}, 
Peter V. Mason\supit{a}, 
Hien T. Nguyen\supit{e}, 
Nicolas Ponthieu\supit{l},
Clem Pryke\supit{m}, 
Steffen Richter\supit{a},\\
Graca M. Rocha\supit{e}, 
Yuki D. Takahashi\supit{i},
Ki Won Yoon\supit{n}.
\skiplinehalf
\supit{a}{California Institute of Technology, Pasadena, USA;} 
\supit{b}{University of Wales, Cardiff, CF243YB, Wales, UK;} 
\supit{c}{Joint ALMA Office, Chile;} 
\supit{d}{University of California, San Diego, USA;} 
\supit{e}{Jet Propulsion Laboratory, Pasadena, USA;} 
\supit{f}{Princeton University, Princeton, NJ, USA;} 
\supit{g}{Commissariat \`{a} l'\'{E}nergie Atomique, Grenoble, France;} 
\supit{h}{Institut d'Astrophysique de Paris, France;} 
\supit{i}{University of California, Berkeley, USA;} 
\supit{j}{Harvard University, USA;} 
\supit{k}{Stanford University, Palo Alto, USA;} 
\supit{l}{Universite Paris XI, Orsay, France;} 
\supit{m}{University of Chicago, USA;} 
\supit{n}{National Institute of Standards and Technology, Boulder, USA.}
}

\authorinfo{(Send correspondence to T. Matsumura)\\T. Matsumura: E-mail: tm@astro.caltech.edu, Telephone: +1 626-395-2147.}

\begin{document}
\maketitle
\vspace{-0.25cm}
\begin{abstract}
We present a method of cross-calibrating the polarization angle of a polarimeter using \bicep\ Galactic observations. \bicep\ was a ground based experiment using an array of 49 pairs of polarization sensitive bolometers observing from the geographic South Pole at 100 and 150~GHz. The \bicep\ polarimeter is calibrated to $\pm 0.01$ in cross-polarization and less than $\pm 0.7^\circ$ in absolute polarization orientation. \bicep\ observed the temperature and polarization of the Galactic plane ($R.A= 100\deg \sim 270\deg$ and $Dec.=-67\deg \sim-48\deg$). We show that the statistical error in the 100~GHz \bicep\ Galaxy map can constrain the polarization angle offset of \wmap\ W~band to $0.6^\circ\pm1.4 ^\circ$. The expected 1$\sigma$ errors on the polarization angle cross-calibration for \planck\ or \epic\ are $1.3\deg$ and $0.3\deg$ at 100 and 150 GHz, respectively. We also discuss the expected improvement of the \bicep\ Galactic field observations with forthcoming \biceptwo\ and \keck\ observations.
\end{abstract}


\vspace{-0.25cm}
\section{Introduction}\label{sec:intro}
The polarization of the cosmic microwave background radiation (CMB) provides a tool for probing the physics of the early Universe. The CMB polarization field is decomposed into even parity E-mode and odd parity B-mode\cite{Hu_White}. Primordial density perturbations result in only E-mode polarization. The E-mode signal was discovered by \dasi\ and characterized by experiments including \boom, \cbi, \maxipol, \QUAD, \wmap, and \bicep~\cite{dasi_kovac,boom_montroy,cbi_sievers,maxipol_wu,quad_brown,wmap3yr_page,bicep_chiang}. Scientific interest in the CMB community moves toward detection of the B-mode signal, which originates from a primordial inflationary gravitational wave background and weak gravitational lensing\cite{pagano_2009}. Numerous kilo-pixel array experiments, including \ebex, \biceptwo, \keck, \polarbear, \quiet, and \spider, are in operation or under construction  to search for B-mode polarization\cite{ebex_britt,bicep2_ogburn,keck_sheehy,polarbear_arnold,spider_crill,spider_filippini}.

While the sensitivity of an experiment increases by employing a large number of detectors, the requirements for controlling systematic effects becomes also stringent\cite{weiss_report}.
Among systematic effects in the experiment, the polarization angle of the detectors is one of the most important quantities to be calibrated. Any miscalibration of the absolute polarization angle of a polarimeter mixes E-mode to B-mode signals, and therefore produces a false B-mode signal. Furthermore, such mixed E and B-mode signals are correlated and non-zero $EB$ correlation indicates a false detection of the CPT violation or cosmic birefringence\cite{wmap7yr_komatsu,cpt_xia,quad_wu}.

The strategy for calibrating the absolute angle of a polarimeter may differ depending on the size of the telescope and the platform of the observatory, i.e. ground-based, balloon-borne and space-borne. The polarization angle of a small ground-based telescope like \bicep\ can be calibrated nearly end-to-end in the optical chain without replying on a calibration source on the sky but rather with a precisely oriented polarized source in front of the aperture in nominal observing conditions. On the other hand, large telescopes or any balloon- or space-borne telescopes are difficult to calibrate in nominal observing conditions without using a polarized sky signal.

Commonly used polarized sources at millimeter wavelengths in the sky are the Crab nebula (Tau A) and Centaurus A (Cen A). The Crab nebula is a supernova remnant that emits highly polarized radiation. Aumont et al. presented the intensity and polarized signals of the Crab nebula observed by \iram\ at 90~GHz\cite{aumont}. Cen A is an active galactic nucleus and Zemcov et al. reported the measurements of Cen A using the QUaD telescope\cite{quad_zemcov}. The reflection from the rim of the Moon is another source of the polarized calibration at millimeter wavelengths for a detector that has a large dynamic range\cite{dasi_leitch}. \wmap\ presented the measurements of the polarized celestial sources, including the Crab nebula, from 23 to 94~GHz\cite{wmap7yr_weiland}. The measurements of the Crab nebula show a consistent polarization angle with Aumont et al.. \planck\ satellite is planning to use this source to calibrate the polarization angle of LFI and HFI detectors\cite{planck_tauber,planck_rosset}.

While these highly polarized compact sources are widely used  for a polarization calibration, a high signal-to-noise diffuse Galactic polarized signal observed by \bicep\ is another polarized source for the angle calibration on the sky.
\bicep\ was a millimeter-wave bolometric polarimeter that is designed to observe the CMB polarization\cite{bicep_yoon}. \bicep\ employs a refractive telescope with a small 24~cm aperture, simplifying the characterization of the end-to-end performance of the polarimeter's entire optical chain.
While the observation of \bicep\ is concentrated on the sky region that is minimally contaminated by the dust and synchrotron emissions, one-fifth of the observational time is dedicated to the Galactic plane observations. 
With systematic effects well controlled, a high signal-to-noise map of the diffuse polarized signal over the Galactic plane makes a standard calibration source on the sky for ongoing and forthcoming CMB polarization experiments.





In Section~\ref{sec:bicepmap} we discuss the statistical and systematic uncertainties in the \bicep\ polarized Galaxy map. In Section~\ref{sec:methods} we discuss the formalism to cross-calibrate the polarization maps produced by an unknown absolute polarization angle polarimeter using the \bicep\ polarization map. In Section \ref{sec:results}, we apply this recipe to cross-calibrate between the \bicep\ 100 GHz and \wmap\ W~band maps, as well as compute the expected constraint on the polarization offset angle for \planck\ and \epic.  

\vspace{-0.2cm}
\section{BICEP polarized Galaxy map}\label{sec:bicepmap}
\bicep\ was a ground based telescope observing from the geographic South Pole. The polarimeter consists of a two lens refractive telescope with a 24~cm aperture and 49 pairs of polarization sensitive bolometers (PSBs) at 100 GHz and 150 GHz with a corresponding beam widths of 0.93$^\circ$ and 0.60$^\circ$, respectively. A detailed description of the \bicep\ instrument is presented in Yoon et al.\cite{bicep_yoon}. 

\begin{figure*}[t]
  \centering
 \begin{tabular}{c}
   \includegraphics[width=14.cm, angle=0]{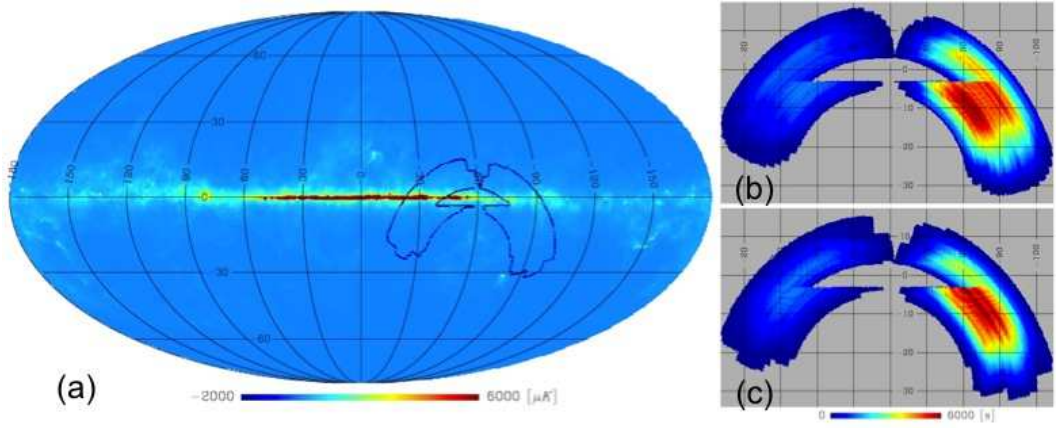}
  \end{tabular}
  \vspace{-0.2cm}
 \caption{\footnotesize \setlength{\baselineskip}{0.95\baselineskip} 
  (a) The two \bicep\ Galactic regions are indicated over the FDS model 8 at 150~GHz\cite{fds}. Right: The integration time per nside=256 of \healpix\ pixel at 100~GHz (b) and 150 GHz (c)\cite{gorski}.}
 \label{fig:Nhits}
\end{figure*}

\bicep\ observed two fields over the Galactic plane as shown in Figure~\ref{fig:Nhits}.  For each field, a telescope scans back-and-forth in azimuth at $2.8^\circ$/s over a $65^\circ$ range at a constant elevation. The elevation is stepped by 0.25$^\circ$ after 50 right and left "half-scans" at a constant elevation. 
The telescope observed with four different orientations about its boresight: $0^\circ, 135^\circ, 180^\circ,$ and $315^\circ$.
Each observation of the single field has a fixed boresight angle and four observations cover all the boresight angles to increase the crosslinking coverage. 


\begin{figure*}[t]
  \centering
 \begin{tabular}{c}
  \includegraphics[width=16.cm, angle=0]{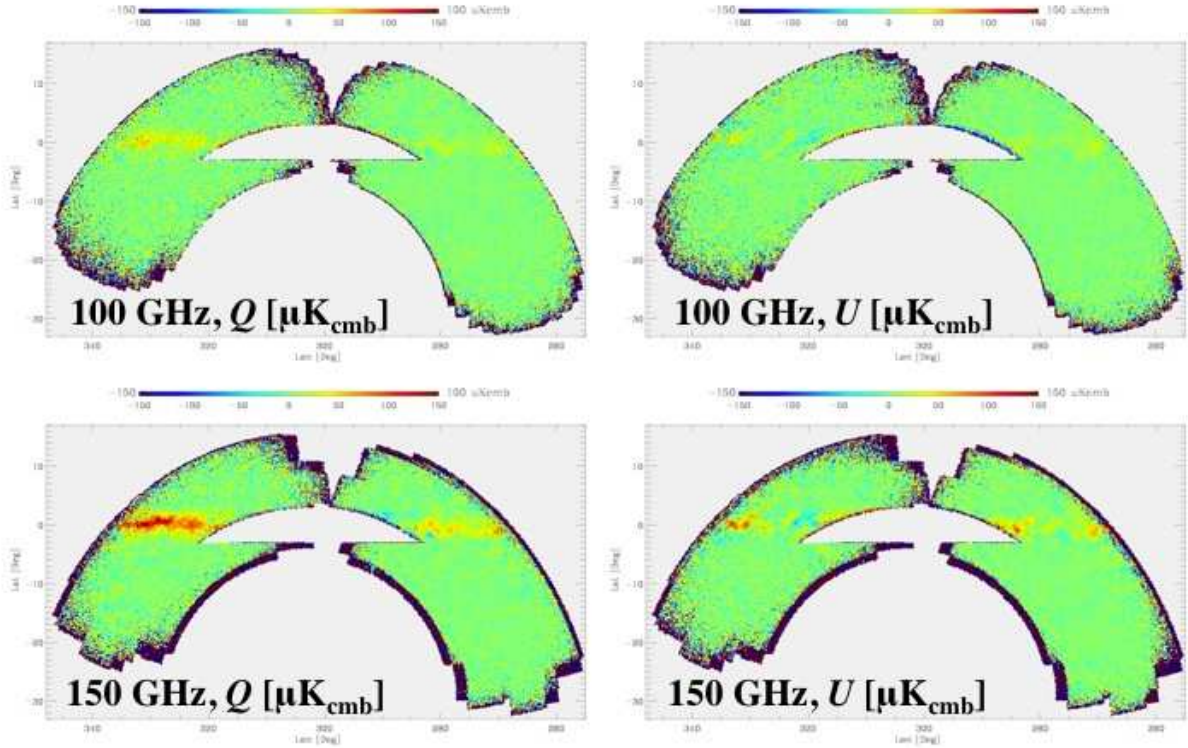}
  \end{tabular}
  \vspace{-0.25cm}
 \caption{\footnotesize \setlength{\baselineskip}{0.95\baselineskip} 
  The $Q$ and $U$ maps of the Galactic fields are shown in the unit of $\mu K_{cmb}$. $Q$ and $U$ are defined in the Galactic coordinate with IAU convention.}
 \label{fig:Galaxy_polmap}
\end{figure*}

\vspace{-0.2cm}
\subsection{Map making}
Figures~\ref{fig:Galaxy_polmap} shows the $Q$ and $U$ maps observed by \bicep\ at 100 and 150~GHz.
This section describes polarized map making, focusing on processes unique to the Galactic field analysis.
The map making process that is common to the CMB analysis is described in Chiang et al.\cite{bicep_chiang}. 

The low-level time stream cleaning is applied to the raw time stream in following steps, (i) deconvolution of a bolometer transfer function, (ii) low-pass filtering at 5~Hz, and (iii) downsampling to 10~Hz.
The $j^{th}$ sample in a half-scan of a gain adjusted pair-differenced time stream is
\begin{equation}
  d_{i}(t_j) = \frac{1}{2}(\frac{d_{i_{\rm A}}(t_j)}{g_{i_{\rm A}}} - \frac{d_{i_{\rm B}}(t_j)}{g_{i_{\rm B}}}),
\end{equation}
where $d_{i_{\rm A,B}}$ are the individual low-level processed time streams of $i^{th}$ PSB pair, and $g_{\rm i_{A,B}}$ are the gain factors for each PSB calibrated at every elevation step.
Common mode noise between the PSB pair, such as thermal fluctuations of the instrument and atmospheric fluctuations, is removed by differencing between the two PSB pair time streams, and we fit a third order polynomial, $F_{i}(t_j)$, to the pair-differenced time stream in order to remove residual $1/f$ noise below $\sim0.1$~Hz. 

When the telescope sweeps over the Galactic plane the variations in the time stream due to the Galactic signal and $1/f$ noise are degenerate. To prevent removing the Galactic signal we apply a mask within the Galactic latitude of $\pm 3\deg$. The mask is applied in the time domain and we only fit the polynomial to the pair-differenced time stream where the mask is not applied. The fitted polynomial is subtracted from the time stream $d_{ij}$ for all the samples in the half-scan. 

In some cases one end of the half-scan lies inside of the Galactic mask. We did not include such half-scan in the map making because the polynomial inside of the mask needs to be extrapolated from the edge of the mask and the extrapolated polynomial does not represent the $1/f$ noise inside of the mask. The half-scans whose two ends lie in the mask are also excluded. Consequently, the recovered maps become ``pac-man" shaped. The recovered region shrinks as the mask width increases. 

We follow the formalism described by Jones et al.\cite{jones_pol} and used for the \bicep\ CMB analysis described in Chiang et al.\cite{bicep_chiang}. The $Q$ and $U$ values of the pixel in the direction on the sky $\vec{p}$ are computed from $d^\prime_{ij} = d_{i}(t_j) - F_{i}(t_j)$ as 
\begin{eqnarray}
  \left [\begin{array}{c}
      Q(\vec{p}) \\   U(\vec{p}) \\ 
    \end{array}\right ] = M \sum_{i}^{N_{PSB}}\sum_{j\subset\vec{p}} w_{ij}
  \left [\begin{array}{c}
      d^\prime_{ij} \alpha \\   d^\prime_{ij}  \beta \\ 
    \end{array}\right ],
  \label{eq:pixelrel}
\end{eqnarray}
where
\begin{eqnarray}
  M^{-1} = \frac{1}{2} \sum_{i}^{N_{PSB}}\sum_{j\subset\vec{p}} w_{ij}  
  \left (\begin{array}{cc}
      \alpha^2_{ij} & \alpha_{ij}\beta_{ij} \\ 
      \alpha_{ij} \beta_{ij} & \beta^2_{ij}  
    \end{array}\right )
\end{eqnarray}
and 
\begin{eqnarray}
  \alpha_{ij} = \gamma_{i_{\rm A}} \cos{2\psi_{i_{\rm A} j}} - \gamma_{i_{\rm B}} \cos{2\psi_{i_{\rm B} j}} \\
  \beta_{ij} = \gamma_{i_{\rm A}} \sin{2\psi_{i_{\rm A} j}} - \gamma_{i_{\rm B}} \sin{2\psi_{i_{\rm B} j}}. 
\end{eqnarray}
The weight $w_{ij}$ is an inverse of a variance of $d^\prime_{ij}$ calculated from the samples outside of the mask in each half-scan. The angle $\psi$ is the PSB orientations projected on the sky and $\gamma = \frac{1-\epsilon}{1+\epsilon}$ is the polarization efficiency factor, where $\epsilon$ is a cross-polarization response. Subscripts $i_{\rm A}$ and $i_{\rm B}$ refers to the $A$ and $B$ bolometers of the $i$th pair. We use nside=256 of \healpix\ pixelization to project $Q$ and $U$ on the sky\cite{gorski}.

\subsection{Pixel noise in the map}\label{subsec:statistical}
Statistical errors in the \bicep\ maps are estimated using jackknife maps. We split the data into three pairs of halves, (i) right and left half-scans, (ii) ($0^{\circ}, 315^{\circ}$) and ($135^{\circ}, 180^{\circ}$) boresight angles, (iii) two sets of detector pairs located in alternating sectors of the 6-sector circular focal plane. We compute the $Q$ and $U$ maps of each data set as $(m_{Q_1}, m_{U_1})$ and $(m_{Q_2}, m_{U_2})$. We compute the difference as
\begin{equation}
  \delta m_Q = \frac{1}{2} (m_{Q_1} - m_{Q_2})\sqrt{N},  \ \ \ \ \  \delta m_U = \frac{1}{2} (m_{U_1} - m_{U_2}) \sqrt{N},
\end{equation}
where $N$ is the number of observations at each \healpix\ pixel. We compute the histogram of $NEQ=\delta m_Q/\sqrt{2f_s}$ and  $NEU=\delta m_U/\sqrt{2f_s}$ (where $f_s = 10$~Hz sampling rate) from the map pixels that meet the criteria of (1) $N > 2000$, (2) the Galactic latitude $|\theta_{glat}| < 3^\circ $ and (3) not being at the edge of the observed regions. (Hereafter we call the region of the sky that meets these criteria a $selected$ region.) We fit the histogram with a Gaussian $A \exp{-\frac{(\delta m-\bar{m})^2}{2 \sigma_0^2}}$. 

\begin{figure*}[t]
  \centering
 \begin{tabular}{c}
     \includegraphics[height=6cm]{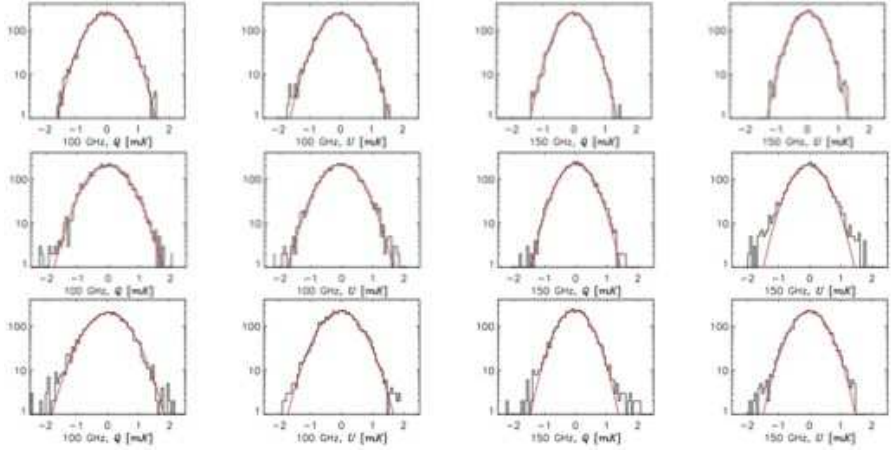}
  \end{tabular}
 \caption{\footnotesize \setlength{\baselineskip}{0.95\baselineskip} 
  The histograms of $\delta m_Q $ and $\delta m_U$ for 100 and 150 GHz are shown for right/left (top), boresight angle (middle), and detector sets (bottom) jackknives. }
  \label{fig:bicepNoiseHist.ps}
\end{figure*}

Figure~\ref{fig:bicepNoiseHist.ps} shows the histograms of the jackknife maps. The noise property is well described by the Gaussian distribution. The averaged $NEQ$ and $NEU$ from the three jackknives are 523 and 507$\muK\sqrt{s}$ for 100~GHz, and 428 and 431$\muK\sqrt{s}$ for 150~GHz, respectively. 
Among the three jackknives the worst $NEQ$ is 16~\% larger than the best one that is from the right and left half-scan jackknife.

\vspace{-0.2cm}
\subsection{Systematic error}\label{subsec:systematics}
The polarization properties of a polarimeter are described by two quantities, PSB orientation $\psi$ and the cross-polarization response $\epsilon$. The PSB orientation $\psi$ is the angle at which the PSB is sensitive to the linear polarization. The cross-polarization response is the response of the PSB to the orthogonally polarized incident radiation.

Two calibration methods are used to measure the cross-polarization. One uses a modulated linearly polarized broadband noise source mounted 200~m away from the \bicep\ telescope. \bicep\ observed the source by raster scans with 18 different boresight rotation angles. The other method uses a rotating wire grid mounted at the cryostat window. The signal is generated by chopping between an ambient absorber and the sky.  The cross-polarization response is measured to within $\pm0.01$.

The PSB orientation is measured with a rotating dielectric sheet mounted in front of the cryostat window in addition to the two methods described to measure the cross-polarization. The measurements were repeated through each observing year and the uncertainty of the individual PSB orientation is 0.1$^\circ$~rms. After the cryostat was opened between 2006 and 2007 observing years, the PSB orientation measurements showed an average of 1$^\circ$ rotation in the absolute polarization angle. Thus, the absolute PSB orientation uncertainty is assigned to be less than 0.7$^\circ$~rms for three years of the observation periods.
The detailed discussion of the polarization calibration of the \bicep\ polarimeter is described in Takahashi et al.\cite{bicep_takahashi}. 

\begin{figure}[t]
  \begin{tabular}{cc}
    \begin{minipage}{0.45\hsize}
      \begin{center}
        \begin{tabular}{cc}
        \includegraphics[height=4.cm]{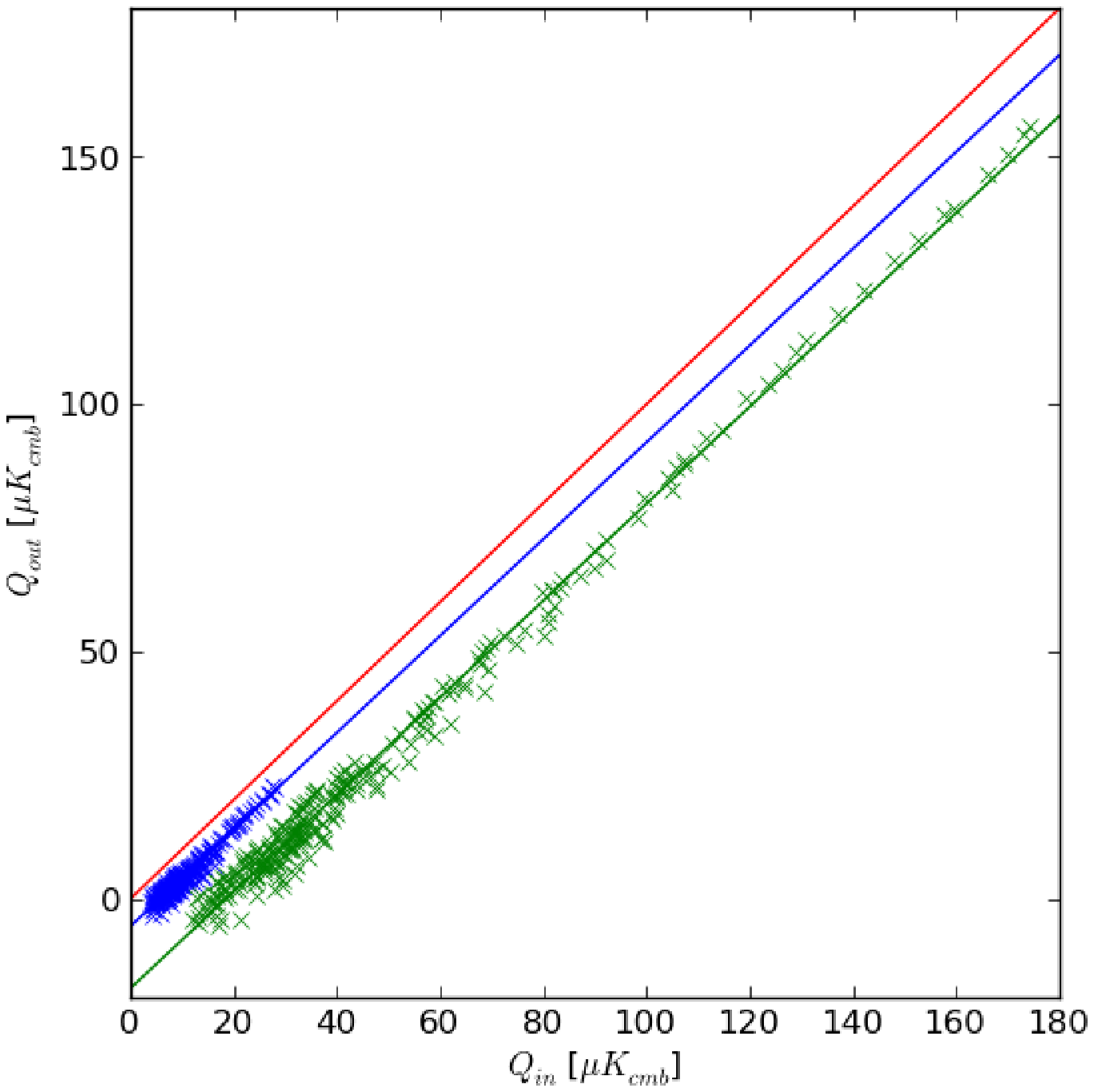}
        \includegraphics[height=4.cm]{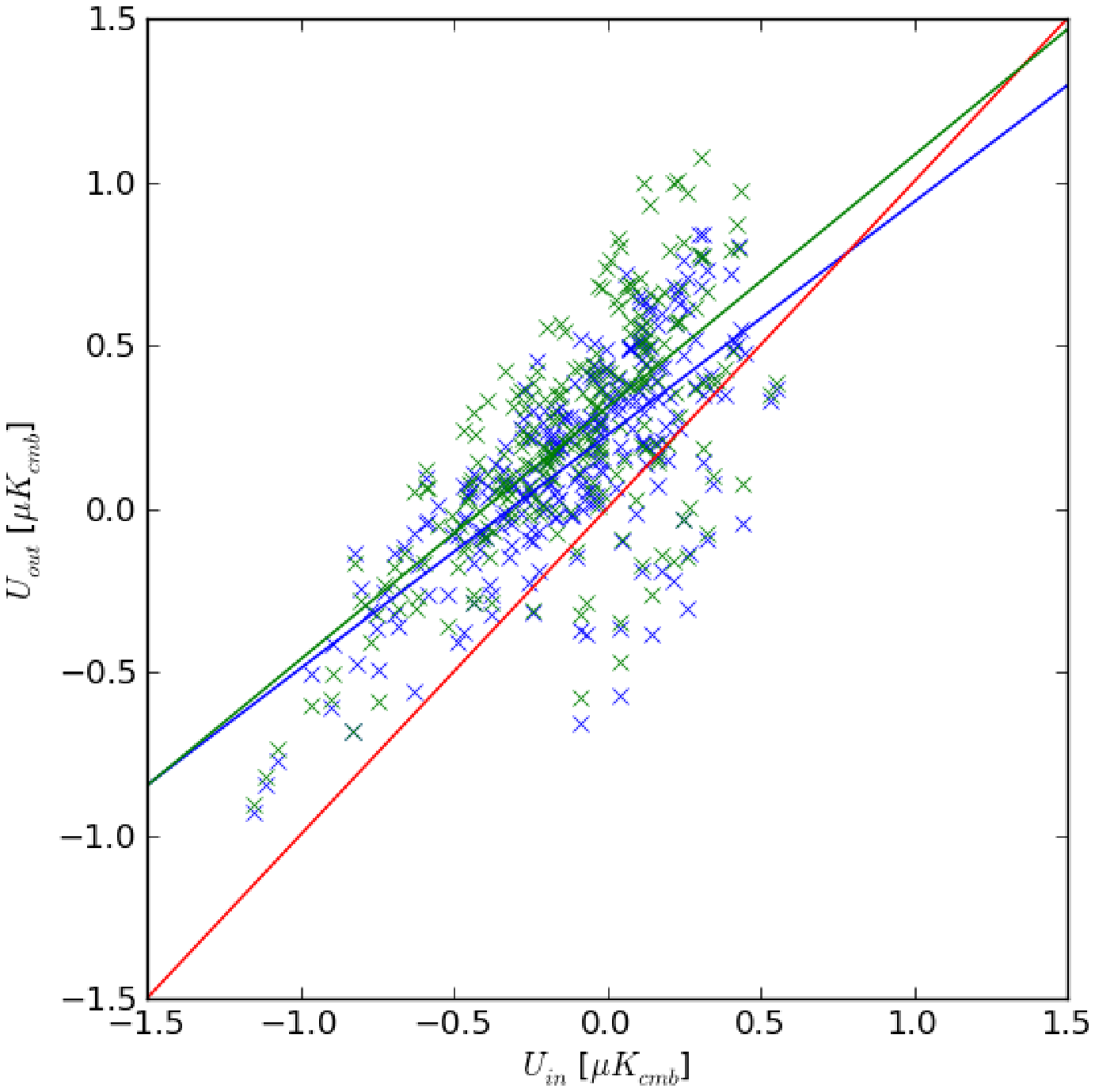}
        \end{tabular}
        \caption{\footnotesize \setlength{\baselineskip}{0.95\baselineskip}Comparison of the filtered $Q$ (and $U$) values of the simulated map against to the input unfiltered $Q$ (and $U$) values. The two linear lines are fits to the 100~GHz (blue) and 150 GHz (green) data. The red line indicates the line that has a slope of 1 with zero offset. The offsets of 100 and 150~GHz $Q$ signals are -5.6 and -18.1, respectively.}
        \label{fig:FilterOffset}
      \end{center}
    \end{minipage}
    \hspace{0.3cm}
   \begin{minipage}{0.45\hsize}
      \begin{center}
        \begin{tabular}{cc}
        \includegraphics[height=4.cm]{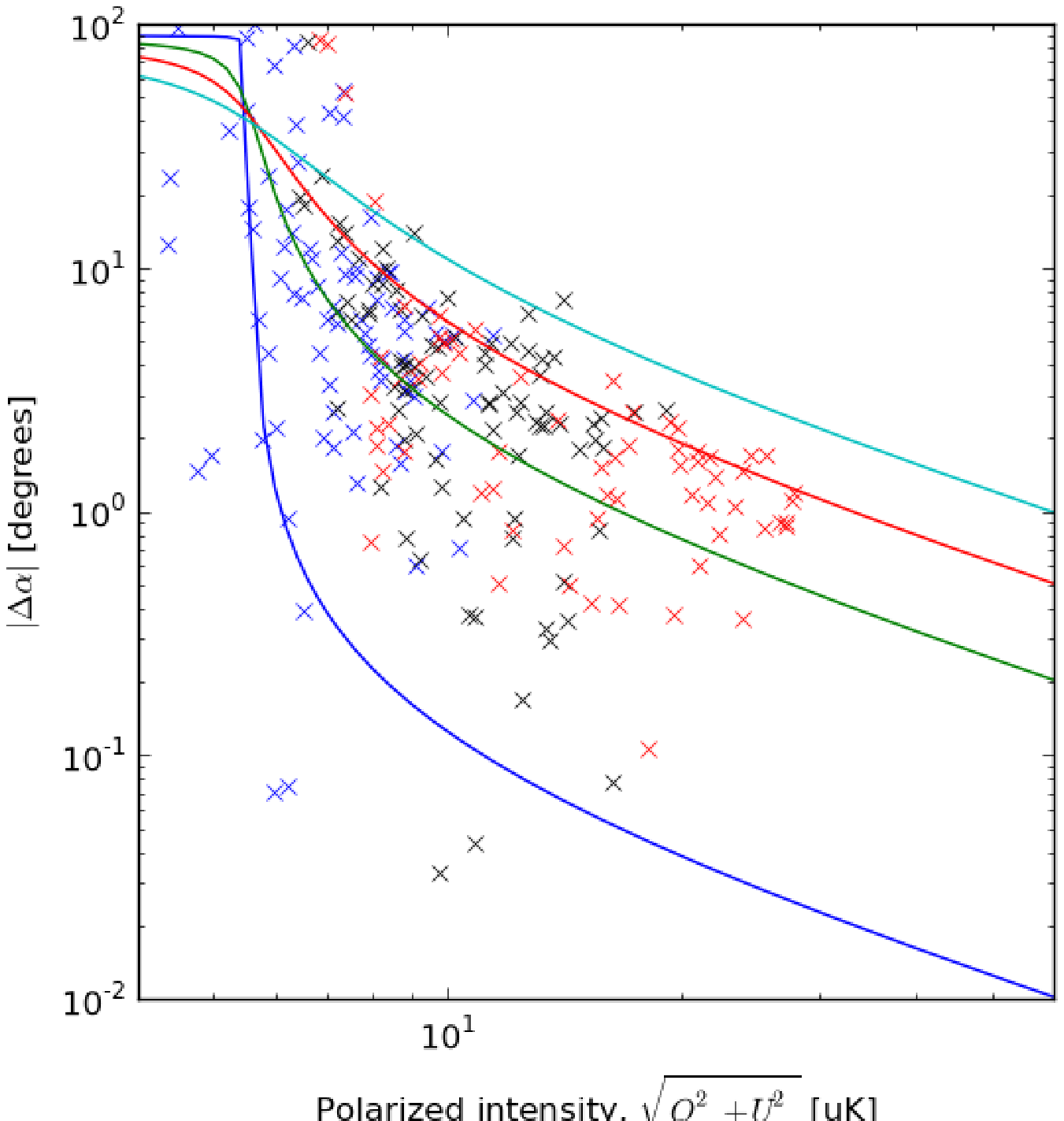}
        \includegraphics[height=4.cm]{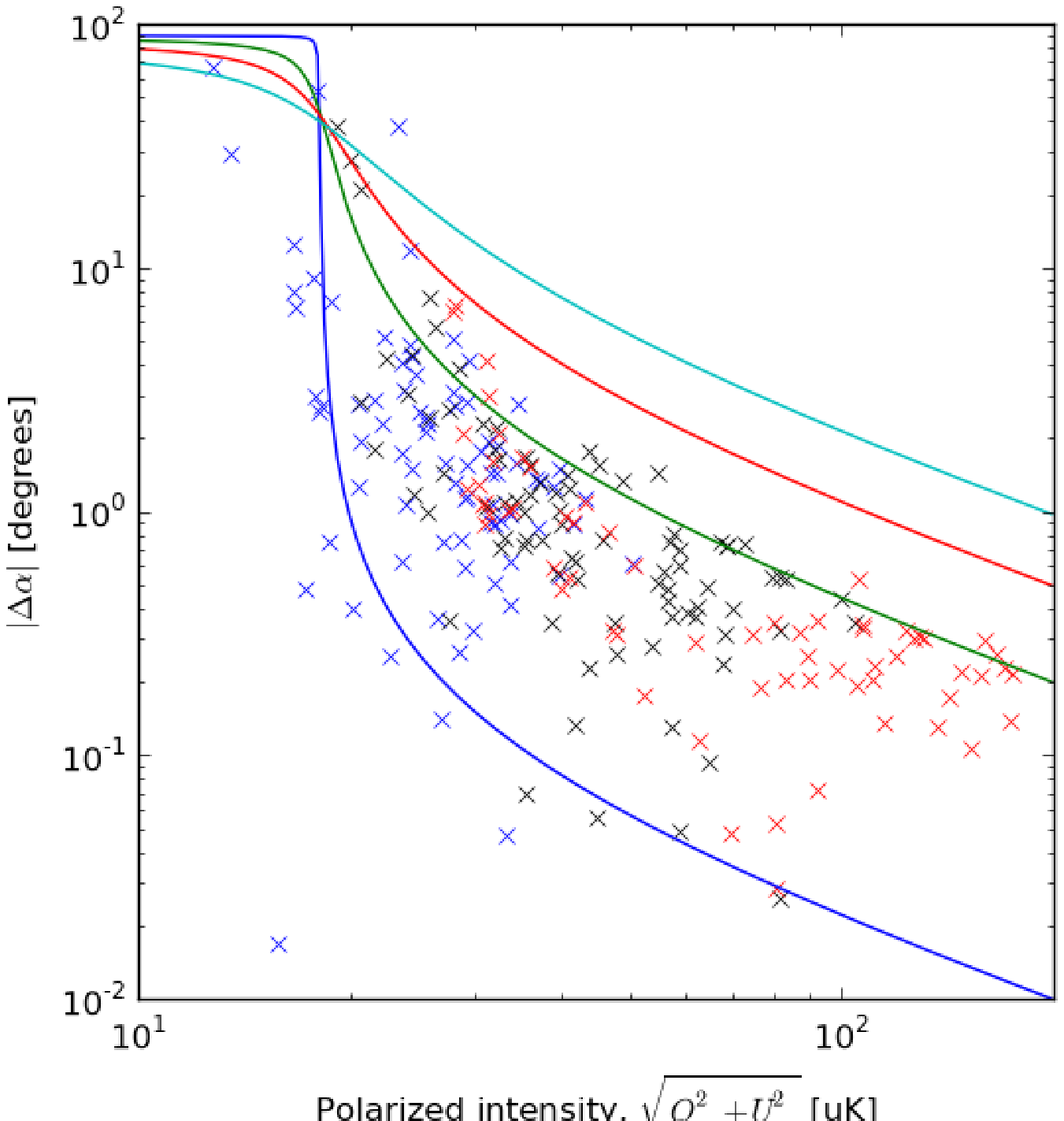}
        \end{tabular}
        \caption{\footnotesize \setlength{\baselineskip}{0.95\baselineskip}The difference of the polarization angle at each pixel before and after applying the BICEP time domain filtering to the simulated maps is plotted. The points are selected from the pixels for the Galactic latitude of $|\theta|<1$ (red), $|\theta|<2$ and $|\theta|>1$ (black), $|\theta|<3$ and $|\theta|>2$ (blue). The four lines are $|\Delta\alpha|$ from Equation~\ref{eq:filteringeffect} with the cases for $\alpha_{in}$ of 0.1$^\circ$, 2$^\circ$, 5$^\circ$ and 10$^\circ$.}
        \label{fig:AngleBias}
      \end{center}
    \end{minipage}
 \end{tabular}
\end{figure}

\vspace{-0.2cm}
\subsection{Effects due to time domain filtering}\label{subsec:filtering}
The subtraction of a polynomial fit from the pair-differenced time stream effectively acts as a high-pass filter in the time domain.  While the purpose of the high-pass filtering is to remove $1/f$ noise, this filtering also removes some modes of the signal.

In order to quantify the amount of the time domain filtering of the Galactic signal, we prepare a simulated polarization map (\healpix\ pixelization of nside=256) that consists of the sum of CMB and FDS maps at 100 and 150 GHz with a beam size of $0.93^{\circ}$ and $0.60^{\circ}$, respectively\cite{fds}. The CMB map is generated by \synfast\ using the cosmological parameters of the standard $\Lambda$CDM model presented in Komatsu et al.\cite{wmap7yr_komatsu,gorski}. The $Q$ polarization of the FDS map is made based on the relationships of $Q/T = c_0 (T/T_{max})^{c_1}$ observed by \bicep\cite{bicep_bierman}. We have used $(c_0,c_1) = (0.007, -0.47)$ and $(0.017, -0.29)$ for 100 and 150~GHz, respectively. 

According to this polarization model, the Galactic $Q$ depends on temperature signal. The FDS model 8 does not have the same level of the emission as it is observed by the BICEP. In order to simulate the realistic level of the Galactic emission we use the temperature $T = \beta T_{FDS}$, where $\beta = 1.30$ and $0.87$ for 100 and 150~GHz, respectively. The $U$ polarization of the FDS is set to be zero for all the pixels.  The simulated maps are smoothed to the beam size of 0.93$^\circ$ and 0.60$^\circ$ for 100 and 150~GHz, respectively.

We generate time ordered data using these simulated maps with the BICEP pointing and apply the same map making as we apply to the real data. Figure~\ref{fig:FilterOffset} shows the correlation between the input and filtered $Q$ and $U$ for the pixels inside of the selected sky region. The relationship of $Q$ before and after applying the time domain filtering is well described by a simple linear relationship. The offset generally depends on the amount of the signal contained at the Galactic plane and the offset is higher when the Galactic signal is higher. This is because the signal level at the mask boundary is significant as compared to the $1/f$ noise, and therefore the interpolated polynomial inside of the mask follows the trend of the Galactic signal instead of the trend from the $1/f$ noise. On the other hand, the $U$ polarization does not show any clear trend. This is because $U$ polarization do not contain any Galactic signal but only the polarization of the CMB. Therefore, there is no characteristic signal increase at the Galactic plane.

Figure~\ref{fig:AngleBias} shows the change of the polarization angle after time domain filtering. The change of the polarization angle $|\Delta \alpha|$ is modeled as
\begin{eqnarray}
  Q_{filt} &=& Q_{in} - Q_0 \\
  U_{filt} &=& U_{in} \\
  \Delta \alpha &=& \frac{1}{2} (\arctan{\frac{U_{filt}}{Q_{filt}}} - \arctan{\frac{U_{in}}{Q_{in}}})  \nonumber \\
  &=& \frac{1}{2} (\arctan{\frac{I_p \sin{2\alpha_{in}}}{I_p \cos{2\alpha_{in}}-Q_0}} - \alpha_{in}),
\label{eq:filteringeffect}
\end{eqnarray}
where $Q_{in}=I_p \cos{2\alpha_{in}}$, $U_{in}=I_p \sin{2\alpha_{in}}$ and $I_p = \sqrt{Q_{in}^2+U_{in}^2}$.  $Q_0$ is the offset to account for the filtering effect.
The time domain filtering effect to the polarization angle ranges from 0.1 to 100 degrees. Therefore, when the offset angle between the \bicep\ map and the map from other experiment is cross-calibrated it is important to apply the same time domain filtering to the other map.

\vspace{-0.3cm}
\section{Estimation of the offset angle and its error}
\label{sec:methods}
We describe a method to detect the overall polarization angle offset between the two polarization maps.
We have two sets of $Q$ and $U$ maps. Ones are the \bicep\ maps as the calibrated maps. The others are maps to be calibrated. In the case of comparing the maps from two different experiments, they do not necessarily have the same beam sizes, and therefore we need to deconvolve the original beam and smooth the two maps to the same beam size. The choice of the beam smoothing varies depending on the beam sizes of \bicep\ and other experiment to be calibrated. In this section we assume that the two sets of maps have a same beam size and are pixelized such that the noise among pixels are not correlated. We discuss the treatment of the different beam size between the separate experiments as a case-by-case basis in Section~\ref{sec:results}.

We write the second and third components of the Stokes parameter of $i$th pixel in the two sets of maps as
\begin{eqnarray}
\mbox{\bicep\ map}&:& \ \ \  (Q_{iB} \pm \delta Q_{iB}, U_{iB} \pm \delta U_{iB}), \\
\mbox{Uncalibrated map}&:& \ \ \  (Q_{i} \pm \delta Q_{i}, U_{i} \pm \delta U_{i}),
\end{eqnarray}
where $\delta Q$ and $\delta U$ indicate the statistical noise. 
We assume that the parent distribution of the pixel noise is a Gaussian described by the standard deviation of $\sigma_{Q_{i(B)}}$, $\sigma_{U_{i(B)}}$, $\sigma_{QU_{i(B)}}$ with a mean of zero.

We relate $Q$ and $U$ of the same pixel on the sky between two experiments by two parameters, offset angle $\delta \alpha$ and the ratio of the polarized amplitudes $\rho$ as 

\begin{eqnarray}
  \left [\begin{array}{c}
      Q_{i} \\   U_{i} \\ 
    \end{array}\right ] = \rho_i
  \left [\begin{array}{cc}
      \cos{2\delta\alpha_i} & \sin{2\delta\alpha_i} \\ 
      -\sin{2\delta\alpha_i} & \cos{2\delta\alpha_i}  \\
    \end{array}\right ]
  \left [\begin{array}{c}
      Q_{B i} \\   U_{B i} \\ 
    \end{array}\right ].
\label{eq:pixelrel}
\end{eqnarray}
We can solve Equation~\ref{eq:pixelrel} for $\rho_i$ and $\alpha_i$ as, 
\begin{eqnarray}
  \left [\begin{array}{c}
      \rho_i \cos{2\delta\alpha_i} \\  \rho_i \sin{2\delta\alpha_i} \\ 
    \end{array}\right ] &=& \frac{1}{Q_{B i}^2+U_{B i}^2}
  \left [\begin{array}{cc}
      Q_{B i} & U_{B i} \\ 
      U_{B i} & -Q_{B i}  \\
    \end{array}\right ]
  \left [\begin{array}{c}
      Q_{i} \\   U_{i} \\ 
    \end{array}\right ], 
\end{eqnarray}
and thus
\begin{eqnarray}
 \label{eq:alpha_i}
  \delta\alpha_i &=& \frac{1}{2}\arctan{\frac{U_{B i} Q_{i}-Q_{B i} U_{i}}{Q_{B i} Q_{i} + U_{B i} U_{i}}}  \nonumber \\
  &=& \alpha_{B i} - \alpha_i, \\
  \label{eq:rho_i}
  \rho_i &=& \sqrt{\frac{Q_{i}^2 + U_{i}^2}{Q_{B i}^2+U_{B i}^2}},
\end{eqnarray}
where $\alpha_{B i} = \frac{1}{2}\arctan{U_{B i}/Q_{B i}}$ and $\alpha_i = \frac{1}{2}\arctan{U_i/Q_i}$.
When the $Q$ and $U$ maps from two separate experiments are identical, $\alpha_i = 0$ and $\rho_i = 1$.

While the polarization calibration can be done in terms of $Q$ and $U$, we express $Q$ and $U$ of two maps in terms of $\delta \alpha$ and $\rho$. This choice was made to mitigate the effect due to the spectral dependence of the instrumental bandpass location and shape mismatch between the two separate experiments.
We discuss the spectral dependence of the polarization angle in Section~\ref{sec:discussion}.

\begin{figure}[t]
  \begin{tabular}{cc}
    \begin{minipage}{0.45\hsize}
      \begin{center}
        \begin{tabular}{c}
          \includegraphics[height=6.0cm]{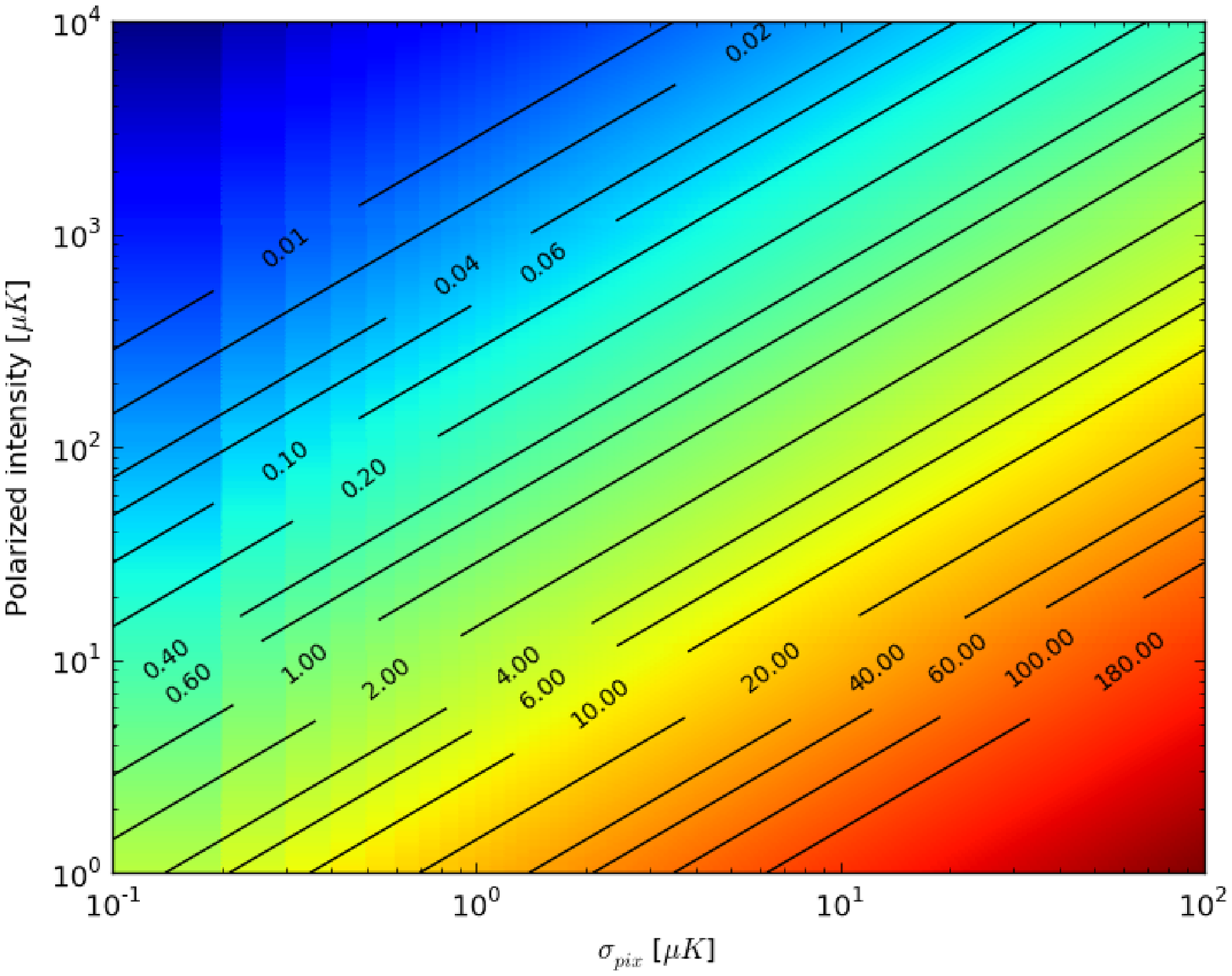}
        \end{tabular}
        \caption{\footnotesize \setlength{\baselineskip}{0.95\baselineskip}This plot shows the angle uncertainty of the polarization signal at a given pixel as a function of a pixel noise and a polarized intensity of the signal. For an example, a map that has a pixel noise of $10~\mu K$ and polarized intensity of $100~\mu K$ has a polarization angle error of 3$^\circ$.}
        \label{fig:2D_SNR}
      \end{center}
    \end{minipage}
    \hspace{0.3cm}
    \begin{minipage}{0.45\hsize}
      \begin{center}
        \begin{tabular}{c}
          \includegraphics[height=6cm]{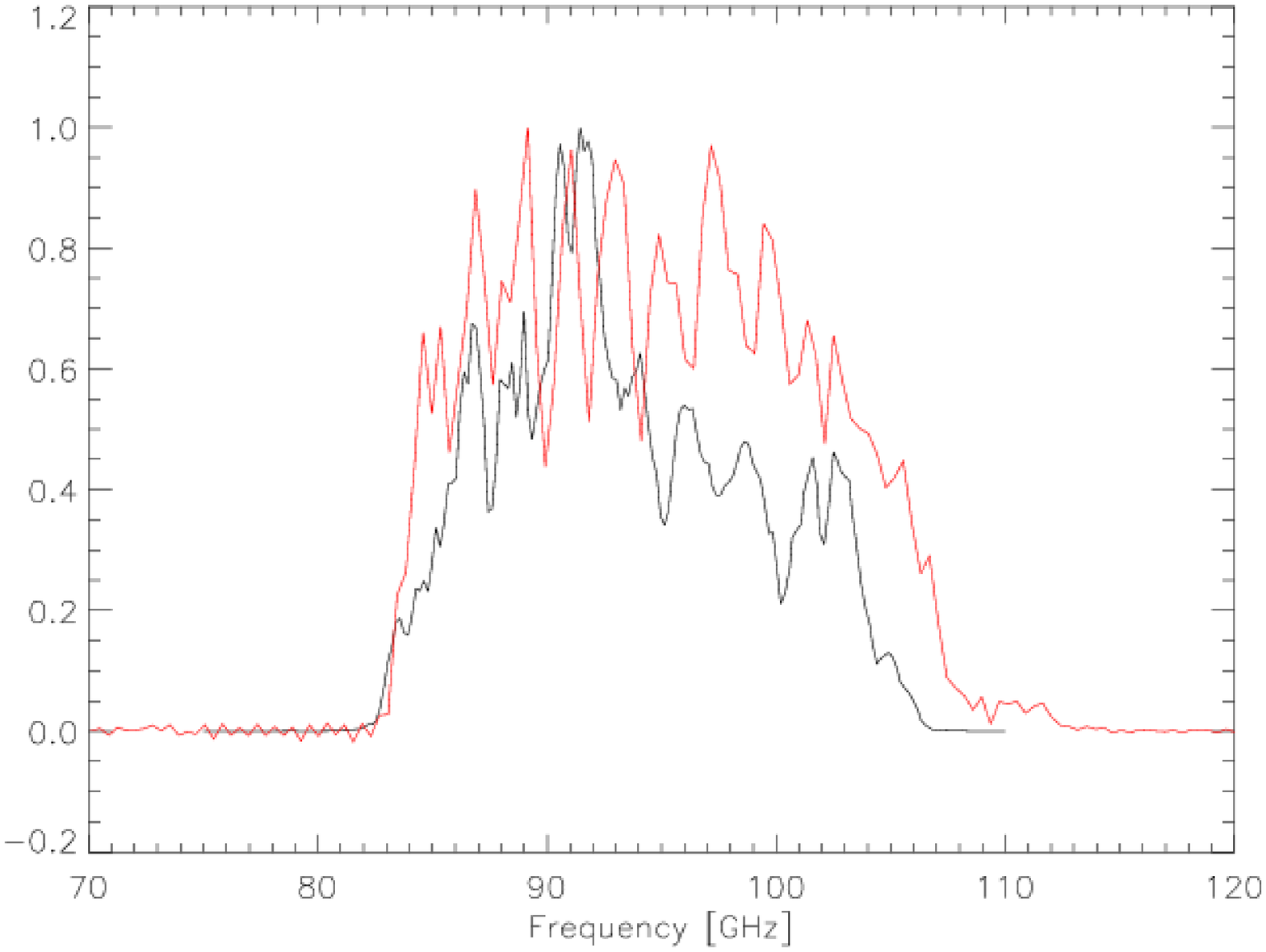}
        \end{tabular}
        \caption{\footnotesize \setlength{\baselineskip}{0.95\baselineskip} The spectra of BICEP (red) and WMAP W~band (black). Both spectra are normalized to the maximum value of 1. The BICEP spectrum is the average of a PSB pair. The WMAP spectrum is the average of the W~band spectra.}
        \label{fig:WMAPvsBICEP}
      \end{center}
    \end{minipage}
  \end{tabular}
\end{figure}

When the $Q$ and $U$ maps contain only signals, we have a perfect knowledge of the offset angle $\delta\alpha$ for every pixel. When the noise is present in the maps, the noise in the map has to be propagated to an error in the offset angle. The error of the offset angle in each pixel $i$ is
\begin{equation}
 \sigma_{\delta\alpha_i}^2 \simeq (\frac{\partial \delta\alpha_i}{\partial Q_{B i}})^2 \sigma_{Q_{B i}}^2 + (\frac{\partial \delta\alpha_i}{\partial U_{B i}})^2 \sigma_{U_{B i}}^2 + \frac{\partial \delta\alpha_i }{\partial Q_{B i} }\frac{\partial \delta\alpha_i}{\partial U_{B i}} \sigma_{QU_{B i}}^2 + (\frac{\partial \delta\alpha_i}{\partial Q_i})^2 \sigma_{Q_i}^2 + (\frac{\partial \delta\alpha_i}{\partial U_i})^2 \sigma_{U_i}^2 + \frac{\partial \delta\alpha_i }{\partial Q_i}\frac{\partial \delta\alpha_i}{\partial U_i} \sigma_{QU_i}^2, 
\end{equation}
where $\sigma_{Q_{B i}}$, $\sigma_{U_{B i}}$, $\sigma_{QU_{B i}}$, $\sigma_{Q_i}$, $\sigma_{U_i}$, $\sigma_{QU_i}$ are the pixel noise associated with $Q_{B i}$ and $U_{B i}$, and $Q_i$ and $U_i$, respectively. The derivative terms are 
\begin{eqnarray}
  \frac{\partial \delta\alpha_i}{\partial Q_{B i}} = -\frac{1}{2}\frac{U_{B i}}{Q_{B i}^2+U_{B i}^2},&&   \frac{\partial \delta\alpha_i}{\partial U_{B i}} = \frac{1}{2}\frac{Q_{B i}}{Q_{B i}^2+U_{B i}^2} \\
  \frac{\partial \delta\alpha_i}{\partial Q_i} = \frac{1}{2}\frac{U_i}{Q_i^2+U_i^2},&&  \frac{\partial \delta\alpha}{\partial U_i} = -\frac{1}{2}\frac{Q_i}{Q_i^2+U_i^2}.
\end{eqnarray}
The derivative terms are inversely proportional to the square of the polarized intensity. This indicates that the error of the offset angle is smaller when the polarized intensity is stronger. Figure~\ref{fig:2D_SNR} shows the angle uncertainty as a function of the pixel noise in $Q$ and $U$ maps and the polarized intensity, $\sqrt{Q^2+U^2}$. 

While a polarization angle $\alpha_i$ varies from $-90$ to $90$ degrees based on the signal and the pixel noise at the given point on the sky, we assume that the distribution of the differenced angle $\delta \alpha_i$ is a Gaussian. We validate this assumption in Section~\ref{sec:results} when we apply this formalism between \bicep\ and \wmap.

The Galaxy is not a single point source, and therefore the estimation of the polarization offset angle improves by including all the available pixels in the map. In order to calculate the mean polarization offset angle, $\delta\alpha_0$, between the reference and the uncalibrated polarimeter maps and the corresponding uncertainty of the mean, we compute the likelihood of $\delta\alpha_0$ as
\begin{equation}
  L \propto \mbox{e}^{-\frac{1}{2}\chi^2(\delta\alpha_0)}, 
   \label{eq:likelihood}
\end{equation}
where
\begin{equation}
  \chi^2(\delta\alpha_0) = \sum_{i} \frac{(\delta\alpha_i - \delta\alpha_0)^2}{ \sigma_{\delta\alpha_i}^{2}} .
\end{equation}


\vspace{-0.4cm}
\section{Results}
\label{sec:results}
We apply the method described in the previous section to \bicep\ and \wmap. We also compute the expected constraint to \planck\ and \epic, by using BICEP Galactic map.

\subsection{Polarization angle offset between \bicep\ and \wmap}
\label{subsec:wmap}
\wmap\ has been observing the temperature and polarization over the full sky\cite{wmap7yr_jarosik}. The spectral bandwidth of \bicep\ 100 GHz and the \wmap\ W~band overlap as shown in Figure~\ref{fig:WMAPvsBICEP}. In this exercise, we assume that the absolute polarization angle of the \wmap\ polarimeter is unknown and we constrain the overall offset angle of the \wmap\ polarization maps using the \bicep\ Galactic map as a polarization calibration source.

Before we apply the formalism described in Section~\ref{sec:methods}, we need to correct the beam size difference between the two experiments. The FWHM of the \bicep\ beam size at 100 GHz is $0.93^\circ$. Each of the $Q$ and $U$ maps of the four \wmap\ W~band differencing assembly is deconvolved with the corresponding \wmap\ $B_l$ and convolved with FWHM of  $0.93^\circ$ Gaussian beam in nside=512 pixelization. We compute the weighted averaged map from the four differencing assembly maps in W~band. The weights are the inverse of the pixel variance of each differencing assembly. We apply the \bicep\ time domain filtering to the averaged \wmap\ W~band map. The filtered $Q$ and $U$ maps are downsampled to $0.92^\circ$ pixel size (nside=64) maps in order to decorrelate the noise among pixels. The \bicep\ $Q$ and $U$ maps are also downsampled to the same pixelization.

The pixel noise of the \wmap\ maps is computed by $\sigma=\frac{\sigma_0}{\sqrt{N_{hits}}}$ where $\sigma_0$ for the \wmap\ W~band differencing assemblies is $(\sigma_{W_1},\sigma_{W_2},\sigma_{W_3},\sigma_{W_4}) = (5.940, 6.612, 6.983, 6.840)$~mK with nside=512 pixelization. We neglect the correlated noise between $Q$ and $U$ for both experiments. The pixel noise of the \bicep\ maps is computed based on the $NEQ(U)$ derived from the right and left jackknife maps.

Once the two sets of the maps and weights are computed in the same pixelization, we impose the criteria to select the pixels. We choose the pixels that meet the criteria of $|\theta_{glat}| < 3^{\circ}$, $N_{hits}$ of \bicep\ $> 2000$ and pixels of which its neighbor do not have $N_{hits}=0$. The second criterion assures that the most of the edge pixels of the map are not included. The second criterion does not exclude the pixel around $282^{\circ}<\phi_{glon}<322^{\circ}$ and $|\theta_{glat}| < 3^{\circ}$ where the edge of the map is not tapered by $N_{hits}$. Therefore, we include the third criterion to exclude all the edge pixels in the maps.

\begin{figure}[t]
    \begin{center}
        \begin{tabular}{c}
        \includegraphics[height=2.7cm]{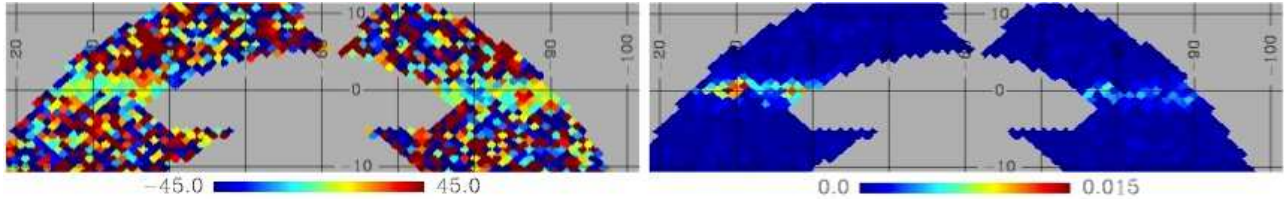}
       \end{tabular}
        \caption{\footnotesize \setlength{\baselineskip}{0.95\baselineskip}The maps of $\delta\alpha_i$ (left) in unit of degrees and weight $=1/\sigma^2_{\delta \alpha_i}$ (right) in unit of degree$^{-2}$. The maps are downsampled to \healpix\ resolution of nside=64. The edge pixels are removed, and therefore the shape of the map does not coincide with the ones in Figure~\ref{fig:Galaxy_polmap}. }
        \label{fig:Angle_weight}
     \end{center}
\end{figure}

\begin{figure}[t]
  \begin{tabular}{c}
    \begin{minipage}{0.45\hsize}
      \begin{center}
        \begin{tabular}{c}
        \includegraphics[height=8cm]{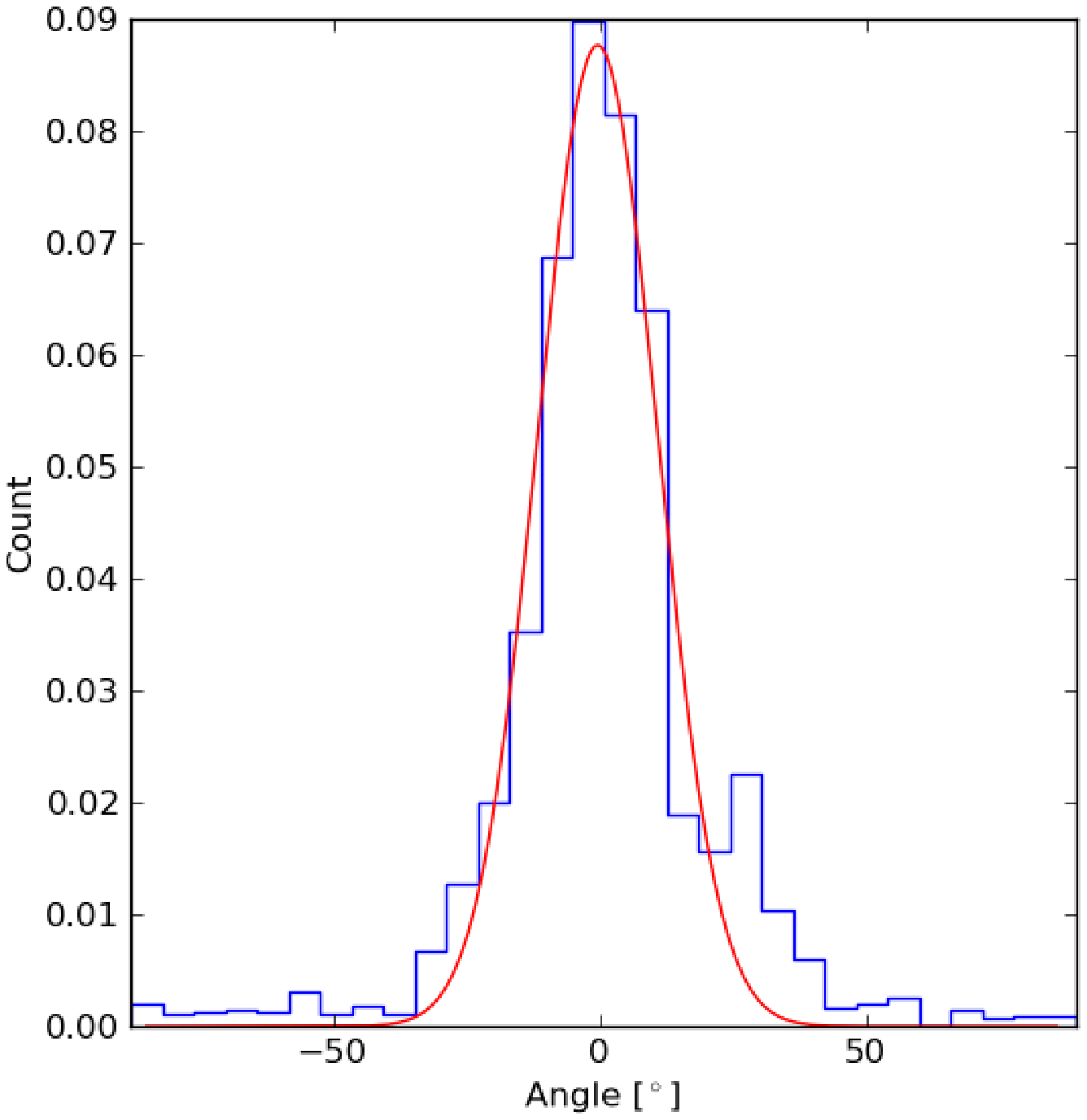}
       \end{tabular}
        \caption{\footnotesize \setlength{\baselineskip}{0.95\baselineskip}The weighted histogram of $\delta \alpha$ between the polarization maps of \bicep\ and \wmap\ and the Gaussian fit are shown. The distribution is well described as a Gaussian.}
        \label{fig:weight_hist}
      \end{center}
    \end{minipage}
    \hspace{0.3cm}
   \begin{minipage}{0.45\hsize}
      \begin{center}
        \begin{tabular}{c}
        \includegraphics[height=8cm]{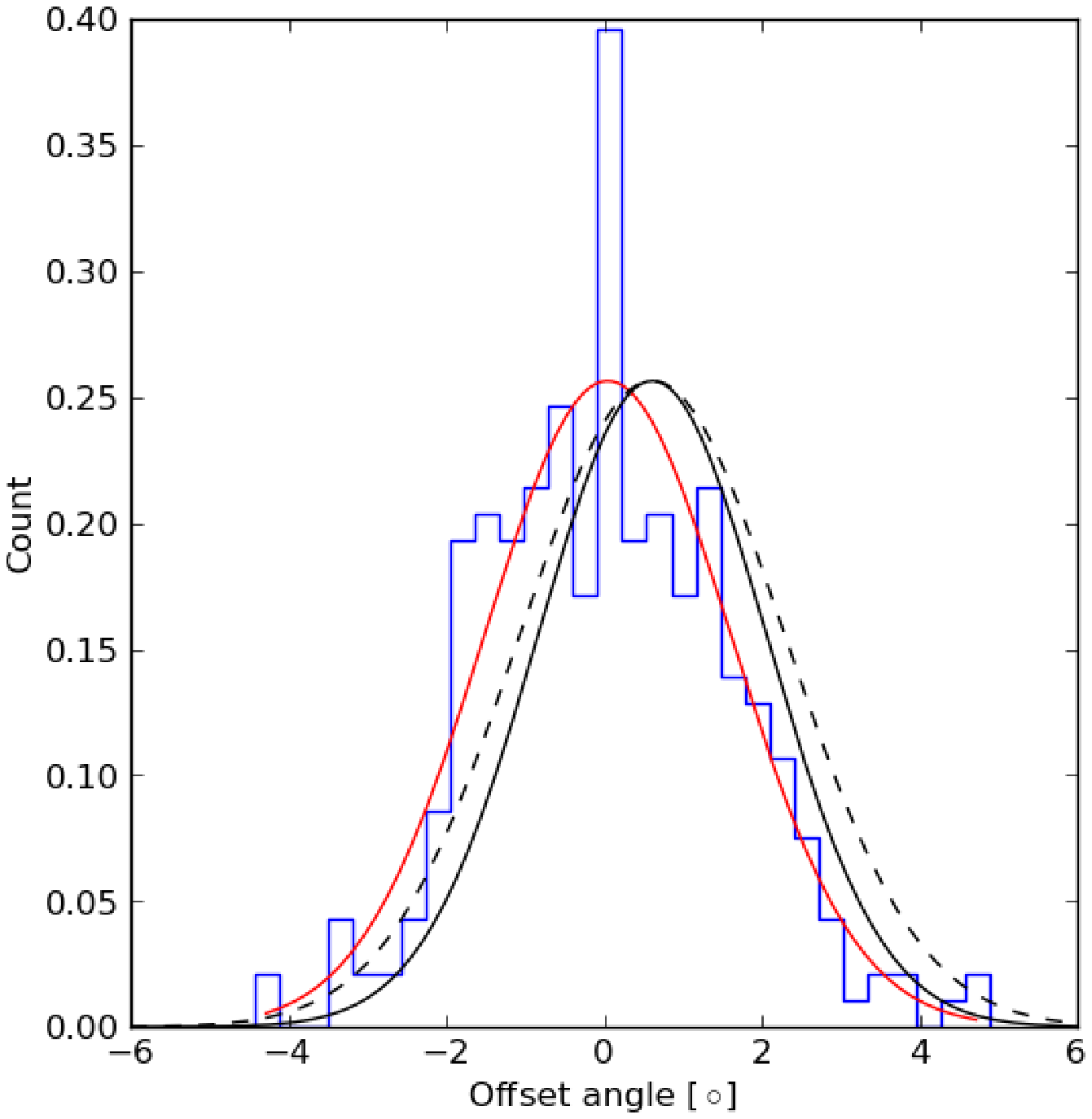}
       \end{tabular}
        \caption{\footnotesize \setlength{\baselineskip}{0.95\baselineskip}The likelihood of $\delta \alpha$ (solid black) with the pixel noise estimated from the right and left jackknife and  $\delta \alpha$ (dash black) with the pixel noise increased by $16~\%$ as a worst pixel noise estimation. The histogram is a mean of the Gaussian fit to $\delta \alpha$ from the two sets of the simulated signal (CMB+FDS) at 100~GHz and the 300 noise realizations. The red curve is the Gaussian fit to the histogram.}
        \label{fig:LikelihoodWMAPvsBICEP}
      \end{center}
    \end{minipage}
 \end{tabular}
\end{figure}

Figure~\ref{fig:Angle_weight} shows the map of offset angle $\delta\alpha_i$ and the weight $1/\sigma^2_{\delta \alpha_i}$. It is clear that the the offset angle is close to zero and the weight is higher at the Galactic plane. 
Figure~\ref{fig:weight_hist} shows the weighted histogram of $\delta\alpha_i$. The mean and the standard deviation of the angle uncertainty of each pixel is $-0.41^\circ$ and $11.2^\circ$, respectively. The distribution of the histogram is well described as a Gaussian distribution.

Figure~\ref{fig:LikelihoodWMAPvsBICEP} shows the likelihood of the offset angle calculated based on Equation~\ref{eq:likelihood} using the \bicep\ and the filtered \wmap\ maps. The black line shows that the mean and the sigma are the 0.6$^\circ$ and $1.4^\circ$ respectively. The dashed line with the same mean has 16\% larger sigma as a worst case pixel noise. 

The histogram in Figure~\ref{fig:LikelihoodWMAPvsBICEP} is the results of the signal and noise simulations. We prepare two sets of maps by adding the white noise of the \bicep\ 100~GHz and \wmap\ W~band to the simulated signal only maps at 100~GHz described in Section~\ref{subsec:filtering}. We repeat computing the mean of the Gaussian fit to the histogram of $\delta\alpha$ from the two sets of the map for the 300 noise realization. The fit to this histogram in Figure~\ref{fig:WMAPvsBICEP} is consistent with the likelihood obtained by the using Equation~\ref{eq:likelihood}.

\begin{table}[t]
  \begin{center}
  \begin{tabular}{c|cc}
      \hline
                & \multicolumn{2}{c}{$1\sigma$ error [$^\circ$]}      \\
      Reference $\times$ Uncalibrated          & $100$~GHz  &  $150$~GHz     \\
      \hline
      \hline
      \bicep\ $\times$ No noise experiment & $1.24$ & $0.27$  \\
      \hline
      \bicep\ $\times$ \wmap\ W-band & $1.45$ &  \\
      \hline
      \bicep\ $\times$ \planck & $1.26$ & $0.27$  \\
      \bicep\ $\times$ \epic-IM (4K option) & $1.24$ & $0.26$ \\
      \hline
      (\bicep, \biceptwo) $\times$ \planck & $1.26$ &  $0.08$\\
      (\bicep, \biceptwo, \keck) $\times$ \planck & $0.23$ &  $0.06$
    \label{tab:FutureExp}
    \end{tabular} 
  \caption{\footnotesize \setlength{\baselineskip}{0.95\baselineskip}The $1\sigma$ statistical error of the polarization angle offset for various combinations of the experiments. The first row, \bicep\ $\times$ No noise experiment, indicates the angle error only due to the \bicep\ statistical noise. \biceptwo\ only has 150~GHz band, and therefore the error in 100~GHz does not show any improvement.}
  \end{center}
\end{table}

\vspace{-0.25cm}
\subsection{Polarization offset angle between BICEP and future experiments}\label{subsec:futureexp}
Any ongoing and forthcoming CMB polarization experiments which observe the \bicep\ Galaxy region can cross-calibrate their polarization angle using the \bicep\ map. As examples, we compute the expected angle constraint for two cases, \bicep\ and \planck, and \bicep\ and \epic\cite{epic-im}.

Table~1 shows the list of $1\sigma$ statistical error from the likelihood in Equation~\ref{eq:likelihood} for 100 GHz and 150 GHz for the two experiments. In this comparison, we assume that the bandpass shape of the two separate experiments is the same and the knowledge of the beam shape is perfect. 

The expected pixel noise of \planck\ and \epic-IM are from \planck\ bluebook and Bock et al., respectively\cite{planck_bluebook,epic-im}. It is clear that the estimate of the $1\sigma$ error of the offset angle between \bicep\ and \wmap\ improves with \bicep\ and \planck\ or \epic-IM. This is because the noise contribution from \planck\ and \epic-IM is much smaller than the case from \wmap\ while the \bicep\ noise stays the same. On the other hand, there is negligible improvement from \planck\ to \epic-IM because the source of the noise in these two cases is limited by the pixel noise of the \bicep\ map.

While the observations of \bicep\ were completed, the ongoing \biceptwo\ and forthcoming \keck\ will improve the sensitivity to the angle calibration. If we assume that \biceptwo\ and \keck\ will spend the same observational time with the same detector sensitivity on the BICEP Galactic field, the expected reduction of the pixel noise is simply scaled by $\sqrt{\frac{N_{Bicep}}{N_{Bicep} +N_{0}}}$, where $N_{Bicep}$ is the number of detectors of \bicep\ and $N_{0}$ is of \biceptwo\ or \biceptwo\ and \keck. We assume that $N_{Bicep}$ is $25$ and $24$ for 100 and 150~GHz, and $N_0$ is 256 for \biceptwo\ 150~GHz and $144\times4$ and $256\times2$ for 100 and 150~GHz of \keck, respectively.
The data combining with \biceptwo\ and \keck\ provide the statistical errors of the offset angle smaller than the systematic errors of the \bicep\ polarimeter itself for both 100 and 150~GHz bands.

\vspace{-0.5cm}
\section{Discussion}\label{sec:discussion}
\vspace{-0.2cm}
\subsection{Comparison between the diffuse Galactic source and the Crab nebula as a polarized calibration source}
We compare the Crab nebula and the BICEP Galactic region as a polarized source. The emission mechanism of the Crab nebula at the millimeter wavelength is dominated by the synchrotron emission. Macias-Perez et al. and Weiland et al. reported that the observed flux has a power law of $\propto(\frac{\nu}{40GHz})^{-0.3 \sim -0.35}$ while the degree of polarization stays constant around 7~\% over the millimeter wavelength\cite{crab_macias_perez,wmap7yr_weiland}. On the other hand, the diffuse dust emission at the Galactic plane increases as a function of frequency\cite{bicep_bierman}. Therefore, the signal-to-noise increases as the bandpass location increases. 

The Crab nebula is a point-like source and the Galaxy is a diffuse source. In order to compare the two sources, we compute the integrated polarized flux of the Galactic field as shown in Table~2.  We also show the integrated polarized flux reported in Weiland et al. and Aumont et al.\cite{aumont,wmap7yr_weiland}. The spatial area of the Galactic field is much larger than that of the Crab nebula. Therefore, the integrated polarized flux of the Galactic signal within the BICEP field is larger than that of the Crab nebula. The total pixel noise from the Galactic field is added in quadrature. As a result, the polarized Galactic source at 150~GHz provides the same order of error as compared to the Crab nebula at W band. 

\begin{table}[t]
  \begin{center}
  \begin{tabular}{c|ccc}
      \hline
      Band &  $Q$ [Jy]  &  $U$ [Jy]  &  Angle [$^\circ$]  \\
      \hline
      \hline
      WMAP, Crab nebula & & & \\
      K band & $-27.13 \pm 0.68$ & $-1.40 \pm 0.08$ & $-88.5\pm0.1 $  \\
      Ka band & $-23.72 \pm 0.45$ & $-1.88 \pm 0.12$ & $-87.7\pm0.1 $  \\
      Q band & $-22.03 \pm 0.60$ & $-2.06 \pm 0.14$ & $-87.3\pm0.2 $  \\
      V band & $-19.25 \pm 0.36$ & $-1.52 \pm 0.24$ & $-87.7\pm0.4 $  \\
      W~band & $-16.58 \pm 0.73$ & $-0.75 \pm 0.42$ & $-88.7\pm0.7 $  \\
      all band combined   &  &  & $0.07 $  \\
      \hline
      Aumont et al., Crab nebula & & &  \\
      90~GHz & & & $-88.8\pm0.2 $  \\
      \hline
      BICEP Galaxy & & & \\
      100~GHz & $114.7 \pm 6.0$ & $29.1 \pm	6.2$ & $\pm1.5 $  \\
      150~GHz & $573.4 \pm 12.2 $ & $195.6 \pm 11.2 $ & $\pm0.5$ 
      \label{tab:CrabvsBICEP}
    \end{tabular} 
    \caption{\footnotesize \setlength{\baselineskip}{0.95\baselineskip}The integrated polarized flux of the Crab nebula and the \bicep\ Galactic region is shown. The polarization convention in this paper and \wmap\ are different, and thus the sign of $U$ is changed from the original WMAP paper. The angle error of the \bicep\ Galactic measurements is the quadrature sum of the pixel noise. $^\dagger$The polarization angle seen by $10^\prime$ beam from Aumont et al. The original literature quoted the polarization angle in equatorial coordinates as $\alpha=148.8\deg$.}
  \end{center}
\end{table}

\vspace{-0.3cm}
\subsection{Effect of spectral mismatch between two experiments}
When the two polarization maps from two separate experiments are cross-calibrated, the spectral bandpass of the two experiments is not necessary the same. We assess the effect of the bandpass mismatch to the offset angle estimation between the \bicep\ 100 GHz band and the \wmap\ W~band. 

Gold et al. derived the synchrotron and dust emission templates by the Markov chain Monte Carlo fitting~\cite{wmap7yr_gold}. We compute the simulated \bicep\ and \wmap\ maps by integrating the sum of the synchrotron and dust template maps over the \bicep\ 100~GHz bandpass and \wmap\ W~band bandpass. We compute the offset angle $\delta\alpha_i$ of each pixel between the two bandpass maps. The median offset angle of all the pixels within the selected sky region is $0.005\deg$. We define the signal-to-noise for each pixel as the ratio of the polarized intensity to the pixel noise. We compute the median and the maximum offset angle of which the pixels are the signal-to-noise $> 3$ are $0.01\deg$ and $0.02\deg$, respectively. 

The \bicep\ 100~GHz map expects a higher contribution of the dust emission as compared to the \wmap\ W~band map because the \bicep\ 100~GHz bandwidth is slightly wider than \wmap\ W~band bandwidth in higher frequency side as shown in Figure~\ref{fig:WMAPvsBICEP}. Gold et al. shows that in the \bicep\ Galactic field the polarization direction of the synchrotron emission is $-26\deg<\alpha_{synch}<0\deg$ and that of the dust emission is $|\alpha_{dust}|<0.3\deg$. Therefore, the overall offset angle between the \bicep\ and \wmap\ maps is expected to show the positive rotation due to the bandpass mismatch. The overall offset angle between the \bicep\ and \wmap\ maps, shown in Figure~\ref{fig:WMAPvsBICEP} , is $0.6\deg\pm1.4\deg$, and the positive mean value is consistent with the bandpass mismatch.

This effect is prominent when the passband of the instrument is located where more than two emission spectra are mixed with nearly the same amplitude. This is because the two sources with different spectral shape can have different polarization angles. 

\vspace{-0.3cm}
\section{Conclusion}\label{sec:conclusion}
We present the polarized diffuse Galactic emissions observed by \bicep\ at 100 and 150~GHz and the method to cross-calibrate the absolute angle between the \bicep\ map and any uncalibrated map. The absolute angle of the \bicep\ polarimeter is calibrated to $\pm0.7~\deg$ and the 1$\sigma$ error of the polarization angle due to the pixel noise of the \bicep\ map is $1.24\deg$ and $0.27\deg$ for 100 and 150~GHz, respectively. 

We apply this method between the \bicep\ and \wmap\ W band maps and cross-calibrate the angle to $0.6\pm1.4\deg$.  The expected 1$\sigma$ errors for the \planck\ 100 and 150~GHz  bands are $1.26\deg$ and $0.27\deg$, respectively. The ongoing and forthcoming \biceptwo\ and \keck\ are expected to reduce the statistical noise of the observations of the \bicep\ Galactic region significantly. 

The \bicep\ Galactic maps provide the polarized Galactic emission as a new angle calibration source for the ongoing and forthcoming CMB B-mode experiments that require the absolute angle calibration to a fraction of a degree. The method of using the Galactic signal as an angle calibration source can be applied to any two experiments if one of the polarimeters is well calibrated. Therefore, when the \planck\ full sky polarization maps are available, the future polarimeters should be able to use the Galactic signal as a calibration source not only with respect to \bicep\ but also to \planck.





\end{document}